\documentclass[preprint,aps]{revtex4}

\begin{document}

\title{Schwarzschild limit of conformal gravity in the presence of macroscopic scalar fields}

\author{Philip D. Mannheim}\email{philip.mannheim@uconn.edu}
\affiliation{Department of Physics,
University of Connecticut, Storrs, CT 06269}

\date{March 6, 2007}

\begin{abstract}

In their original study of conformal gravity, a candidate alternate gravitational  theory, Mannheim and Kazanas showed that in any empty vacuum region exterior to a localized static spherically symmetric gravitational source, the geometry would reduce to the standard attractive gravity Schwarzschild geometry on solar system distance scales. In a recent paper Flanagan has argued that this would not be the case if the source has associated with it a macroscopic scalar field which makes a non-zero contribution to the energy-momentum tensor in the otherwise empty exterior region. In this paper we examine Flanagan's analysis and show that even with such long range scalar fields, the standard Schwarzschild phenomenology is still recovered.

\end{abstract}

\maketitle

\section{Introduction}

As a theory of gravity, conformal gravity is very appealing since it not only possesses the full metric structure associated with the standard Einstein gravitational theory, it in addition also possesses the local conformal symmetry which is characteristic of theories of elementary particles in which particle masses are to be generated entirely via dynamically generated symmetry breaking phase transitions in the vacuum. With the action of the conformal theory being required to be invariant under local conformal changes of the metric of the form $g_{\mu\nu}(x) \rightarrow e^{2\alpha(x)}g_{\mu\nu}(x)$, in the conformal theory the purely gravitational sector of the action is uniquely prescribed to be of the form 
\begin{eqnarray}
I_W&=&-\alpha_g\int d^4x (-g)^{1/2}C_{\lambda\mu\nu\kappa}
C^{\lambda\mu\nu\kappa}
\nonumber \\
&=&-2\alpha_g\int d^4x
(-g)^{1/2}\left[R_{\mu\kappa}R^{\mu\kappa}-\frac{1}{3}
(R^{\alpha}_{\phantom{\alpha}\alpha})^2\right]
\label{1}
\end{eqnarray}
where $C_{\lambda\mu\nu\kappa}$ is the conformal Weyl tensor and $\alpha_g$ is a necessarily  dimensionless gravitational coupling constant. Similarly, with the matter action equally needing to be conformally coupled at the level of the Lagrangian, the prototypical conformal matter action consisting of a fermion field $\psi(x)$ with spin connection $\Gamma_{\mu}(x)$ and a scalar field $S(x)$ is uniquely specified to be of the form 
\begin{equation}
I_M=-\int d^4x(-g)^{1/2}\left[\frac{1}{2}S^{;\mu}
S_{;\mu}-\frac{1}{12}S^2R^\mu_{\phantom         
{\mu}\mu}
+\lambda S^4
+i\bar{\psi}\gamma^{\mu}(x)[\partial_\mu+\Gamma_\mu(x)]             
\psi -hS\bar{\psi}\psi\right],
\label{2}
\end{equation}                                 
where the coefficient of the $S^2R^\mu_{\phantom{\mu}\mu}$ term is uniquely required to be equal to the indicated negative factor of $-1/12$ and where $h$ and $\lambda$ are dimensionless coupling constants. On the theoretical side, giving the scalar field $S(x)$ a non-zero vacuum expectation value will spontaneously break the conformal symmetry and give the fermion a mass (as needed for particle physics), while on the phenomenological side the conformal theory has been found capable of readily solving both the dark matter and dark energy problems which currently challenge the standard theory (see e.g. \cite{Mannheim2006} where full bibliography and background are given). 

For the above $I_W+I_M$ action functional variation with respect to the matter fields yields the equations of motion
\begin{equation}
i \gamma^{\mu}(x)[\partial_{\mu} +\Gamma_\mu(x)]                              
\psi - h S \psi = 0,
\label{3}
\end{equation}                                 
and 
\begin{equation}
S^{;\mu}_{\phantom{\mu};\mu}+\frac{1}{6}SR^\mu_{\phantom{\mu}\mu}
-4\lambda S^3 +h\bar{\psi}\psi=0,
\label{4}
\end{equation}                                 
while functional variation with respect to the metric yields
\begin{equation}
4\alpha_g W^{\mu\nu}=T^{\mu\nu},
\label{5}
\end{equation}
where $W^{\mu\nu}$ is given by
\begin{eqnarray}
W^{\mu \nu}&= &
\frac{1}{2}g^{\mu\nu}(R^{\alpha}_{\phantom{\alpha}\alpha})   
^{;\beta}_{\phantom{;\beta};\beta}+
R^{\mu\nu;\beta}_{\phantom{\mu\nu;\beta};\beta}                     
 -R^{\mu\beta;\nu}_{\phantom{\mu\beta;\nu};\beta}                        
-R^{\nu \beta;\mu}_{\phantom{\nu \beta;\mu};\beta}                          
 - 2R^{\mu\beta}R^{\nu}_{\phantom{\nu}\beta}                                    
+\frac{1}{2}g^{\mu\nu}R_{\alpha\beta}R^{\alpha\beta}
\nonumber \\
&&-\frac{2}{3}g^{\mu\nu}(R^{\alpha}_{\phantom{\alpha}\alpha})          
^{;\beta}_{\phantom{;\beta};\beta}                                              
+\frac{2}{3}(R^{\alpha}_{\phantom{\alpha}\alpha})^{;\mu;\nu}                           
+\frac{2}{3} R^{\alpha}_{\phantom{\alpha}\alpha}
R^{\mu\nu}                              
-\frac{1}{6}g^{\mu\nu}(R^{\alpha}_{\phantom{\alpha}\alpha})^2,
\label{6}
\end{eqnarray}                                 
and $T^{\mu\nu}$ is given by
\begin{eqnarray}
T^{\mu \nu} &=& i \bar{\psi} \gamma^{\mu}(x)[
\partial^{\nu}                    
+\Gamma^\nu(x)]                                                                 
\psi+\frac{2}{3}S^{;\mu} S^{;\nu} 
-\frac{1}{6}g^{\mu\nu}S^{;\alpha} S_{;\alpha}
-\frac{1}{3}SS^{;\mu;\nu}
+\frac{1}{3}g^{\mu\nu}SS^{;\alpha}_{\phantom{;\alpha};\alpha}  
\nonumber \\             
&&                          
-\frac{1}{6}S^2\left(R^{\mu\nu}
-\frac{1}{2}g^{\mu\nu}R^\alpha_{\phantom{\alpha}\alpha}\right)         
-g^{\mu\nu}\lambda S^4. 
\label{7}
\end{eqnarray}                                 
For the purposes of studying localized static spherically symmetric matter sources we can represent the fermionic part $\hat{T}^{\mu\nu}=i \bar{\psi} \gamma^{\mu}(x)[\partial^{\nu} +\Gamma^\nu(x)]\psi$ of the full energy-momentum tensor $T^{\mu\nu}$ as the perfect fluid 
\begin{equation}
\hat{T}^{\mu\nu}=\frac{1}{c}\left[(\rho+p)U^{\mu}U^{\nu}+pg^{\mu\nu}\right].
\label{8}
\end{equation}                                 
With the trace of this $\hat{T}^{\mu\nu}$ being given by
\begin{equation}
\hat{T}^\mu_{\phantom{\mu}\mu}=\frac{1}{c}\left[3p-\rho\right]=i \bar{\psi} \gamma^{\mu}(x)[\partial_{\mu} +\Gamma_\mu(x)]\psi=
hS \bar{\psi} \psi,
\label{9}
\end{equation}                                 
we find that Eq. (\ref{4}) can be rewritten as
\begin{equation}
\frac{1}{6}S^2R^\mu_{\phantom{\mu}\mu}
=\frac{1}{c}\left[\rho-3p\right]+4\lambda S^4 -SS^{;\mu}_{\phantom{\mu};\mu},
\label{10}
\end{equation}                                 
a form that will prove central in the following.

Given the form of $W^{\mu\nu}$ in Eq. (\ref{6}), it can immediately be shown \cite{Mannheim1989} that in any  $\rho(r>R)=0$, $p(r>R)=0$, $S(r>R)=0$ empty, static, spherically symmetric exterior $r>R$ region where the full $T^{\mu\nu}$ vanishes, the $R^{\mu\nu}=0$ Schwarzschild solution is indeed an exterior empty vacuum solution ($W^{\mu\nu}=0$) to the theory; with its matching to the interior region at the surface $r=R$ showing \cite{Mannheim1994} that the coefficient $2\beta$ of the $-1/r$ term in the familiar Schwarzschild solution $-g_{00}=1/g_{rr}=1-2\beta/r$ is indeed the positive one needed for gravitational attraction. If however, we instead allow the scalar field to be non-zero in the $r>R$ region where the localized $\rho(r)$ and $p(r)$ are to vanish, on making the specific local conformal transformation $S(r) \rightarrow e^{-\alpha(r)} S(r)$ which brings the scalar field to a constant value $S_0$ in such $r>R$  exterior regions, and on then, as per the analysis followed in  \cite{Flanagan2006}, dropping the ensuing $-g^{\mu\nu}\lambda S_0^4$ term from Eq. (\ref{7}),  the full $T^{\mu\nu}$ is then reduced to 
\begin{eqnarray}
T^{\mu\nu}(r)&=&\frac{1}{c}\left[(\rho+p)U^{\mu}U^{\nu}+pg^{\mu\nu}\right]-\frac{1}{6}S_0^2\left(R^{\mu\nu}
-\frac{1}{2}g^{\mu\nu}R^\alpha_{\phantom{\alpha}\alpha}\right),
\nonumber \\
T^{\mu\nu}(r>R)&=&-\frac{1}{6}S_0^2\left(R^{\mu\nu}
-\frac{1}{2}g^{\mu\nu}R^\alpha_{\phantom{\alpha}\alpha}\right),
\label{11}
\end{eqnarray}                                 
to now not necessarily vanish outside the source. However, before proceeding to analyze the implications of Eq. (\ref{11}),  we note that the dropping of the $-g^{\mu\nu}\lambda S_0^4$ term while retaining the $-(S_0^2/6)\left(R^{\mu\nu}
-\frac{1}{2}g^{\mu\nu}R^\alpha_{\phantom{\alpha}\alpha}\right)$ term is not actually consistent with the notion of symmetry breaking in curved spacetime,  since the very existence of a constant scalar field which is non-zero throughout the spacetime is a global, long range order effect and not a local one, an effect which thus couples to the global geometry. Indeed, it is the very non-vanishing of the $-g^{\mu\nu}\lambda S_0^4$ term in $T^{\mu\nu}$  which is needed to support the non-zero $S_0$ in the first place, with the $\lambda S^4 -(S^2/12)R^\alpha_{\phantom{\alpha}\alpha}$ term in the action of Eq. (\ref{2}) serving as a double-well potential in spacetimes (such as for instance de Sitter) in which the global curvature is non-zero. The $-(S_0^2/6)\left(R^{\mu\nu}
-\frac{1}{2}g^{\mu\nu}R^\alpha_{\phantom{\alpha}\alpha}\right)$ term in $T^{\mu\nu}$ is thus just as much a part of the cosmological background as the $-g^{\mu\nu}\lambda S_0^4$ term. In fact, as we shall discuss in detail below, the effect of the introduction into the cosmological background of a local mass source is not actually associated with the constancy of the scalar field, but rather with the departure from constancy which is brought about by the very presence of the localized mass source. And it will actually be this spatially dependent departure from constant field and not the $-(S_0^2/12)R^\alpha_{\phantom{\alpha}\alpha}$ term itself which will be responsible for the gravity produced by a localized source. 

Nonetheless, even with this proviso, it is still of interest to determine what particular local gravity would be associated with the truncated energy-momentum tensor given in Eq. (\ref{11}), and to this end we note that in the exterior region an $R^{\mu\nu}=0$ geometry will nonetheless actually cause the $T^{\mu\nu}(r>R)$ of Eq. (\ref{11}) to vanish, with an $R^{\mu\nu}=0$ geometry still being an exact $W^{\mu\nu}=0$, $T^{\mu\nu}=0$ solution to the theory in the $r>R$ region. However, with Eq. (\ref{10}) reducing to
\begin{equation}
\frac{S_0^2}{6}R^\mu_{\phantom{\mu}\mu}=\frac{1}{c}\left[\rho-3p\right],
\label{12}
\end{equation}                                 
and with the standard Einstein equations for a standard pure $\hat{T}^{\mu\nu}$ perfect fluid source and no macroscopic scalar field, viz.
\begin{equation}
-\frac{c^3}{8\pi G}\left(R^{\mu\nu}
-\frac{1}{2}g^{\mu\nu}R^\alpha_{\phantom{\alpha}\alpha}\right)=\frac{1}{c}\left[(\rho+p)U^{\mu}U^{\nu}+pg^{\mu\nu}\right],
\label{13}
\end{equation}                                 
entailing that
\begin{equation}
\frac{c^3}{8\pi G}R^\mu_{\phantom{\mu}\mu}=\frac{1}{c}\left[3p-\rho\right],
\label{14}
\end{equation}                                 
the flip in sign between Eqs. (12) and (14) (as occasioned by the $-1/12$ factor in Eq. (\ref{2})) would suggest that the conformal gravity $R_{\mu\nu}=0$ Schwarzschild solution would correspond to repulsive rather than attractive gravity. Precisely such a concern has been raised by Flanagan \cite{Flanagan2006} and Van Acoleyen \cite{VanAcoleyen2006}, to thus call  into question the viability of the conformal theory. However, an explicit non-perturbative numerical analysis of Eqs. (\ref{3}) - (\ref{5}) by Wood and Moreau \cite{Wood2001} has found that even with a macroscopic scalar field, solar system gravity is still attractive. It is the purpose of this paper to reconcile these differing claims, and in particular to examine the analysis of \cite{Flanagan2006} in detail. However, in order to do this we need first to recall how it was that Mannheim and Kazanas were able to recover standard attractive gravity in an empty exterior region in the first place.

\section{Empty exterior region solution}

To deal with the case of a static spherically symmetric geometry, rather than use a metric in the standard form
\begin{equation}
ds^2 = -b(\rho)c^2dt^2 + a(\rho)d\rho ^2 + \rho ^2 d\Omega_2,
\label{15}
\end{equation}
in the conformal theory it is convenient to make the general coordinate transformation
\begin{equation}
\rho = p(r),~~B(r) = {r^2 b(r)\over
p^2(r)},~~                             
 A(r) = {r^2 a(r) p^{\prime 2}(r)\over p^2 (r)}
\label{16}
\end{equation}
with an initially arbitrary function $p(r)$, as this brings the metric of Eq. (\ref{15}) to the form 
\begin{equation}
ds^2 = {p^2(r)\over r^2}\left[-B(r) c^2dt^2 + A(r) dr^2 + r^2
d\Omega_2\right].
\label{17}
\end{equation}
On now choosing $p(r)$ according to 
\begin{equation}
-{1\over p(r)} = \int{dr\over r^2[a(r)b(r)]^{1/2}},
\label{18}
\end{equation}
the function $A(r)$ is made equal to $1/B(r)$, with the metric of Eq. (\ref{18}) then being found to take the form
\begin{equation}
ds^2= {p^2(r)\over r^2}\left[-B(r)c^2 dt^2+ {dr^2\over B(r)} + r^2
d\Omega_2\right].
\label{19}
\end{equation}
As such, the above sequence of purely kinematic coordinate transformations trades the two metric coefficients $a(\rho)$ and $b(\rho)$ for two other coefficients $B(r)$ and $p(r)$, but does so in way in which one of them, $p(r)$, appears purely as an overall multiplier of the metric. Since the theory is conformal invariant, both of the functions $W^{\mu\nu}$ and $T^{\mu\nu}$ which appear in Eq. (\ref{5}) transform under the conformal transformation $g_{\mu\nu}\rightarrow e^{2\alpha(x)}g_{\mu\nu}$ as $W^{\mu\nu}\rightarrow e^{-6\alpha(x)}W^{\mu\nu}$, $T^{\mu\nu}\rightarrow e^{-6\alpha(x)}T^{\mu\nu}$. The factor $p^2(r)/r^2$ can thus be scaled out of the theory, with the full content of the conformal theory being contained in the metric
\begin{equation}
ds^2= -B(r)c^2 dt^2+ {dr^2\over B(r)} + r^2d\Omega_2,
\label{20}
\end{equation}
a metric which now contains just a single unknown metric coefficient $B(r)$. 

The great utility of the metric of Eq. (\ref{20}) is that for it \cite{Mannheim1994,Mannheim2006} the combination $W^0_{{\phantom 0} 0} - W^r_{{\phantom r} r}$ evaluates exactly and without any approximation whatsoever to the extraordinarily compact form 
\begin{equation}                                                                              
\frac{3}{B}\left(W^0_{{\phantom 0} 0} - W^r_{{\phantom r} r}\right)
=B^{\prime \prime \prime \prime} + \frac{4
B^{\prime \prime \prime}}{r}.
\label{21}
\end{equation}                     
(The primes denote derivatives with respect to $r$.) Consequently, on recognizing that $B^{\prime \prime \prime \prime} + 4B^{\prime \prime \prime}/r$ is the radial piece of the fourth order Laplacian $\nabla^4 B$, and recalling that the tracelessness and covariant conservation of  $W^{\mu\nu}$ [$W^0_{{\phantom 0} 0} +W^r_{{\phantom r} r} +2W^{\theta}_{{\phantom \theta} \theta}=0$, $(B^{\prime}/2B-1/r)(W^0_{{\phantom 0} 0} - W^r_{{\phantom r} r})-(d/dr+4/r)W^r_{{\phantom r} r}=0$] entail that it only has one independent component in the static spherically symmetric case, we find that the equation of motion of Eq. (\ref{5}) can be written very compactly as          
\begin{equation}                                                                               
\nabla^4 B(r) = f(r)
\label{22}
\end{equation}      
where the source function $f(r)$ is defined via                   
\begin{equation}                                                                              
f(r) = \frac{3}{4\alpha_g B}\left(T^0_{{\phantom 0} 0} -
T^r_{{\phantom r} r}\right)= \frac{1}{4\alpha_g }\left[-\frac{3(\rho+p)}{cB}
+SS^{\prime \prime}-2S^{\prime 2}\right]                  
\label{23}
\end{equation}                                                  
as evaluated here with the metric given in  Eq. (\ref{20}) and the explicit form for $T^{\mu\nu}$ given in Eq. (\ref{7}). 

As a differential equation, Eq. (\ref{22}) admits of the general solution 
\begin{equation}
B(r)=-\frac{r}{2}\int_0^r
dr^{\prime}r^{\prime 2}f(r^{\prime})
-\frac{1}{6r}\int_0^r
dr^{\prime}r^{\prime 4}f(r^{\prime})
-\frac{1}{2}\int_r^\infty
dr^{\prime}r^{\prime 3}f(r^{\prime})
-\frac{r^2}{6}\int_r^\infty
dr^{\prime}r^{\prime }f(r^{\prime}),
\label{24}
\end{equation}                                 
together with an additional  cosmologically relevant $w-kr^2$ term which solves the homogeneous
$\nabla^4 B(r)=0$. Thus, in cases in which the entire source $f(r)$ vanishes in the $r>R$ region (viz. the $S(r>R)=0$, $\rho(r>R)=0$, $p(r>R)=0$ case considered in \cite{Mannheim1989,Mannheim1994})  the metric exterior to the source is given as 
\begin{equation}
B(r>R)=1-\frac{2\beta}{r}+\gamma r
\label{25}
\end{equation}                                 
where we have introduced the coefficients
\begin{equation}
\gamma= -\frac{1}{2}\int_0^R
dr^{\prime}r^{\prime 2}f(r^{\prime}),~~
2\beta=\frac{1}{6}\int_0^R
dr^{\prime}r^{\prime 4}f(r^{\prime}).
\label{26}
\end{equation}                                 
With the $\beta$ and $\gamma$ coefficients being associated with different moments of the source they are in principle different, with the successful phenomenological fitting of conformal gravity to galactic rotation curve data \cite{Mannheim2006} without the use of any dark matter leading to the numerical values $\beta=1.48\times 10^5$ cm, $\gamma=5.42\times 10^{-41}$ cm$^{-1}$ per unit solar mass of material. With such values, the linear potential modification to the Newtonian $1/r$ potential is thus negligible on solar system distance scales, with the standard solar system Schwarzschild phenomenology then following. For these particular values of $\beta$ and $\gamma$ we note also that the ratio $(\beta/\gamma)^{1/2}$ is equal to $0.52 \times 10^{23}$ cm, a ratio which is thus galactic in scale rather than stellar. Hence the function $f(r)$ which appears in Eq. (\ref{26}) must be singular, since if it were uniform it would otherwise lead to $(\beta/\gamma)^{1/2} \sim R$ for a star of radius $R$.  Thus unlike the second order Poisson equation which is not sensitive to the structure of the source, precisely because it is  a higher derivative equation, the fourth order Poisson equation does explore the singularity  structure of the source. To get a sense of what the sensitivity to singularities might be, we note that the illustrative (though not mandatory) choice of  source 
\begin{equation}
f(r)=-2p \frac{\delta(r)}{r^2}
-\frac{3q}{2}\left[\nabla^2
-\frac{r^2}{12}\nabla^4\right]\left[\frac{\delta(r)}{r^2}\right]
\label{27}
\end{equation}                                 
leads to
\begin{equation}
2\beta=q,~~\gamma=p.
\label{28}
\end{equation}                                 
With the $p$-dependent term only contributing to the second moment and the $q$-dependent term only contributing to the fourth, the source of Eq. (\ref{27}) shows that in the presence of singularities the second and fourth moments are logically independent, a point that will prove to be of relevance in the discussion  below. (As a source we note that its $q$-dependent part can be written as the $\epsilon \rightarrow 0$ limit of the quantity $6q\epsilon(9r^4-10r^2\epsilon^2-3\epsilon^4)/\pi(r^2+\epsilon^2)^5$, a quantity which (for $q>0$) is positive in $r > O(\epsilon)$ and negative in $r <O(\epsilon)$, but which traps a singularity at $r=0$ in the $\epsilon \rightarrow 0$ limit, causing its net contribution to the second moment integral of the source to vanish. Such singularities while foreign to the conventional standard gravity wisdom are not forbidden by any known gravitational observation -- even the cherished positivity of the energy density is not actually mandated by observation, as the positivity of a moment  integral does not entail the positivity of its integrand.) With the sign of $\beta$ being fixed by the sign of $q$ and with the sign of $q$ itself being fixed by the sign of $\alpha_g$ in Eq. (\ref{23}), there thus will be a choice of sign for the coupling constant $\alpha_g$ for which the coefficient $\beta$ will be positive, with it thus being possible to unambiguously fix the sign of the coefficient of the $1/r$ term in the conformal theory. (In the standard theory this same outcome is achieved by an a priori choice for the sign of $G$.) Additionally we note that the sign of $\alpha_g$ bears no relation to the sign of the coefficient of the $S^2R^\mu_{\phantom         
{\mu}\mu}$ term in the action of Eq. (\ref{2}), a point which will also prove to be of relevance below.

Proceeding next to the case where there is to be a non-vanishing $S(r)$ in the $r>R$ region, we note that with the scalar field transforming as $S(x)\rightarrow e^{-\alpha(x)}S(x)$ under a conformal transformation, one should not expect that the transformation which will bring the metric to the form of Eq. (\ref{20}) will be just the one which will completely cancel the spatial dependence of $S(r)$ and bring it to a constant $S_0$. Consequently, if one takes the metric to be of the form given in Eq. (\ref{20}), one should not expect the associated $T^{\mu\nu}$  to contain no derivatives of $S(r)$. Working with the metric in the form of Eq. (\ref{20}) thus obligates working with the full spatially dependent $S(r)$ in Eq. (\ref{7}). For the particular choice of metric of Eq. (\ref{20}) Eq. (\ref{10}) is found to take the form
\begin{equation}
\frac{1}{6}S^2\left[B^{\prime\prime}+\frac{4B^{\prime}}{r}+\frac{2B}{r^2}-\frac{2}{r^2}\right] =\frac{ (\rho -3p)}{c}
+4\lambda S^4 -S\left[BS^{\prime\prime}+\frac{2BS^{\prime}}{r}+B^{\prime}S^{\prime}\right],
\label{29}
\end{equation}
which when taken in conjunction with  Eq. (\ref{22}) and an equation of state for the perfect fluid thus defines the problem in the non-vanishing $S(r)$ case. Unfortunately no exact analytic solution to this set of coupled equations is currently known, and  for the moment one must resort to a numerical treatment. A numerical solution has been provided in \cite{Wood2001} which finds that the standard attractive Schwarzschild phenomenology continues to hold in the solar system, with the wisdom of the solution of Eq. (\ref{25}) thus being maintained. (Numerically, the effect of the gradient of the scalar field on both the geometry and particle trajectories in the exterior region was found to be within current experimental bounds, while leading to a small departure from the standard Schwarzschild phenomenology which the authors of \cite{Wood2001} suggest could be responsible for the anomalous acceleration of the Pioneer spacecraft -- the radial gradient of any macroscopic scalar field associated with the sun would naturally point toward it.) Armed with the above analysis we turn now to the work of Flanagan.

\section{Linearization in isotropic coordinates} 

With the metric associated the exterior empty vacuum solution of Eq. (\ref{25}) being close to flat in the intermediate region where the radial distance $r$ is neither too small nor too large, in this intermediate region one can thus consider solving for the case in which there is a constant exterior macroscopic scalar field by linearizing the theory around flat spacetime. This then is the approach of Flanagan, and since one would not expect that it would matter what choice of perturbed coordinate system one might choose to work in, Flanagan opted to work with isotropic coordinates rather than standard ones. However, as we shall see below, on working the problem  through in linearized standard coordinates we will actually reach a conclusion quite different from the one that Flanagan reached in his linearized isotropic coordinate study. We shall thus describe both of the linearization calculations and shall then reconcile the difference between them.

For an isotropic coordinate system with metric
\begin{equation}
ds^2 = -H(\rho)c^2dt^2 + J(\rho)\left[d\rho ^2 + \rho ^2 d\Omega_2\right],
\label{30}
\end{equation}
linearization is associated with the metric
\begin{equation}
ds^2 = -[1+h(\rho)]c^2dt^2 + [1+j(\rho)]\left[d\rho ^2 + \rho ^2 d\Omega_2\right]
\label{31}
\end{equation}
where $h(\rho)$ and $j(\rho)$ are small. With the explicit form  of the exact $W^{\mu\nu}$ associated with the metric of Eq. (\ref{30}) being given in \cite{Kazanas1991}, its linearization is found to lead to
\begin{eqnarray}
W^{tt} &=& \frac{1}{3}\left[j^{\prime\prime\prime\prime}-h^{\prime\prime\prime\prime}\right]
+\frac{4}{3\rho}\left[j^{\prime\prime\prime}-h^{\prime\prime\prime}\right] =
\frac{1}{3}\nabla^4[j-h],
\nonumber \\             
W^{\rho\rho}&=&                          
\frac{1}{3\rho}\left[j^{\prime\prime\prime}-h^{\prime\prime\prime}\right]
+\frac{2}{3\rho^2}\left[j^{\prime\prime}-h^{\prime\prime}\right]  
-\frac{2}{3\rho^3}\left[j^{\prime}-h^{\prime}\right] .
\label{32}
\end{eqnarray}                                 
Similarly a linearization of the Einstein tensor $G^{\mu\nu}=R^{\mu\nu}-(1/2)g^{\mu\nu}R^{\alpha}_{\phantom{\alpha}\alpha}$ in isotropic coordinates yields 
\begin{eqnarray}
G^{t}_{\phantom{t}t}&=& -\left[j^{\prime\prime}+\frac{2}{\rho}j^{\prime}\right]=-\nabla^2j,
\nonumber \\             
G^{\rho}_{\phantom{\rho}\rho}&=& -\frac{1}{\rho}\left[j^{\prime}+h^{\prime}\right],
\nonumber \\             
G^{\theta}_{\phantom{\theta}\theta}&=& -\frac{1}{2}\left[j^{\prime\prime}+h^{\prime\prime}\right]
-\frac{1}{2\rho}\left[j^{\prime}+h^{\prime}\right],
\label{33}
\end{eqnarray}                                 
with the Ricci scalar being given by
\begin{equation}
R^{\mu}_{\phantom{\mu}\mu}=\left[2j^{\prime\prime}+h^{\prime\prime}\right]
+\frac{2}{\rho}\left[2j^{\prime}+h^{\prime}\right]=\nabla^2\left[2j+h\right].
\label{34}
\end{equation}
Thus in the non-relativistic perfect fluid limit where $p(\rho) \ll \rho(\rho)$, Eqs. (\ref{5}), (\ref{11}) and (\ref{12}) reduce to
\begin{equation}
4\alpha_gW^{tt}=\frac{4\alpha_g}{3}\nabla^4[j-h]=\frac{ \rho(\rho)}{c} -\frac{S_0^2}{6}\nabla^2j,
\label{35}
\end{equation}
\begin{equation}        
4\alpha_gW^{\rho\rho}=\frac{4\alpha_g}{3}\left[\frac{1}{\rho}(j^{\prime\prime\prime}-h^{\prime\prime\prime})
+\frac{2}{\rho^2}(j^{\prime\prime}-h^{\prime\prime})
-\frac{2}{\rho^3}(j^{\prime}-h^{\prime})\right]= \frac{S_0^2}{6\rho}\left[j^{\prime}+h^{\prime}\right],
\label{36}
\end{equation}
\begin{equation}             
\frac{S_0^2}{6}R^\mu_{\phantom{\mu}\mu}=\frac{S_0^2}{6}\nabla^2\left[2j+h\right]=\frac{\rho(\rho)}{c}.
\label{37}
\end{equation}                                 
In his paper Flanagan explored various classes of solution to this set of equations, but if we are to recover the linearized isotropic coordinate system Schwarzschild solution, viz. 
\begin{equation}
ds^2 = -\left[1-\frac{2MG}{\rho}\right]c^2dt^2 + \left[1+\frac{2MG}{\rho}\right]\left[d\rho ^2 + \rho ^2 d\Omega_2\right],
\label{38}
\end{equation}
we should set $j(\rho)=-h(\rho)$, with Eq. (37) then leading to the wrong sign for the coefficient of the  $1/\rho$ term. This then is the concern raised by Flanagan.

\section{Linearization in standard coordinates}

To assess the significance of this result it is instructive to repeat the analysis in standard coordinates.
For the standard coordinate system with metric
\begin{equation}
ds^2 = -B(r)c^2dt^2 + A(r)dr ^2 +r^2 d\Omega_2,
\label{39}
\end{equation}
 linearization is associated with the metric
\begin{equation}
ds^2 = -[1+b(r)]c^2dt^2 + [1+a(r)]dr^2 + r^2 d\Omega_2.
\label{40}
\end{equation}
This time the linearization of the exact $W^{\mu\nu}$  given in \cite{Kazanas1991} is found to lead to
\begin{eqnarray}
W^{tt} &=& -\frac{1}{3}\left[b^{\prime\prime\prime\prime}+\frac{4b^{\prime\prime\prime}}{r}
+\frac{a^{\prime\prime\prime}}{r} +\frac{a^{\prime\prime}}{r^2}
-\frac{2a^{\prime}}{r^3}+\frac{2a}{r^4}\right],
\nonumber \\             
W^{rr}&=&                          
-\frac{1}{3}\left[\frac{b^{\prime\prime\prime}}{r}+\frac{2b^{\prime\prime}}{r^2}-\frac{2b^{\prime}}{r^3}
+\frac{a^{\prime\prime}}{r^2}
-\frac{2a}{r^4}\right].
\label{41}
\end{eqnarray}                                 
Similarly, a linearization of the Einstein tensor  in standard coordinates yields 
\begin{eqnarray}
G^{t}_{\phantom{t}t}&=& \frac{a^{\prime}}{r}+\frac{a}{r^2},
\nonumber \\             
G^{r}_{\phantom{r}r}&=& -\frac{b^{\prime}}{r}+\frac{a}{r^2},
\nonumber \\             
G^{\theta}_{\phantom{\theta}\theta}&=& -\frac{b^{\prime\prime}}{2}-\frac{b^{\prime}}{2r}+\frac{a^{\prime}}{2r},
\label{42}
\end{eqnarray}                                 
with the Ricci scalar being given by
\begin{equation}
R^{\mu}_{\phantom{\mu}\mu}=b^{\prime\prime}+\frac{2b^{\prime}}{r}
-\frac{2a^{\prime}}{r}-\frac{2a}{r^2}.
\label{43}
\end{equation}
Thus in the non-relativistic perfect fluid limit where $p(r) \ll \rho(r)$, Eqs. (\ref{5}), (\ref{11}) and (\ref{12}) reduce to
\begin{equation}
4\alpha_gW^{tt}=-\frac{4\alpha_g}{3}\left[b^{\prime\prime\prime\prime}+\frac{4b^{\prime\prime\prime}}{r}
+\frac{a^{\prime\prime\prime}}{r} +\frac{a^{\prime\prime}}{r^2}
-\frac{2a^{\prime}}{r^3}+\frac{2a}{r^4}\right] 
= \frac{\rho}{c} +\frac{S_0^2}{6}\left[\frac{a^{\prime}}{r}+\frac{a}{r^2}\right],
\label{44}
\end{equation}                                 
\begin{equation}           
4\alpha_gW^{rr}=-\frac{4\alpha_g}{3}\left[\frac{b^{\prime\prime\prime}}{r}+\frac{2b^{\prime\prime}}{r^2}-\frac{2b^{\prime}}{r^3}+\frac{a^{\prime\prime}}{r^2}-\frac{2a}{r^4}\right]=-\frac{S_0^2}{6}\left[-\frac{b^{\prime}}{r}+\frac{a}{r^2}\right],
\label{45}
\end{equation}                                 
\begin{equation}          
\frac{S_0^2}{6}R^\mu_{\phantom{\mu}\mu}=\frac{S_0^2}{6}\left[b^{\prime\prime}+\frac{2b^{\prime}}{r}
-\frac{2a^{\prime}}{r}-\frac{2a}{r^2}\right]=\frac{\rho}{c},
\label{46}
\end{equation}                                 
to thus define the problem in the standard coordinate system.

Before attempting to look for solutions to Eqs. (\ref{44}) - (\ref{46}) it is useful to recall how the linearized standard coordinate Schwarzschild solution is obtained in the standard Einstein theory. Specifically, one  linearizes Eqs. (\ref{13}) and (\ref{14}) using the metric of Eq. (\ref{40}) and ignores the perfect fluid pressure, to obtain 
\begin{eqnarray}
-\frac{c^3}{8 \pi G}\left[ \frac{a^{\prime}}{r}+\frac{a}{r^2}\right]&=&-\frac{\rho}{c},
\nonumber \\             
\frac{c^3}{8 \pi G}\left[b^{\prime\prime}+\frac{2b^{\prime}}{r}
-\frac{2a^{\prime}}{r}-\frac{2a}{r^2}\right]&=& -\frac{\rho}{c},
\label{47}
\end{eqnarray}                                 
viz.
\begin{equation}
\frac{c^4}{8 \pi G}\left[ \frac{a^{\prime}}{r}+\frac{a}{r^2}\right]=\rho,~~
\frac{c^4}{8 \pi G}\left[b^{\prime\prime}+\frac{2b^{\prime}}{r}\right]= \rho,
\label{48}
\end{equation}                                 
with general all $r$ solution
\begin{eqnarray}
a(r)&=& \frac{8\pi G}{c^4r}\int_0^rdr^{\prime}r^{\prime 2}\rho(r^{\prime}),
\nonumber \\             
b(r)&=&-\frac{8\pi G}{c^4r}\int_0^rdr^{\prime}r^{\prime 2}\rho(r^{\prime}) -\frac{8\pi G}{c^4}\int_r^\infty dr^{\prime}r^{\prime }\rho(r^{\prime}),
\label{49}
\end{eqnarray}                                 
and familiar $r>R$ exterior solution
\begin{equation}          
b(r>R)=-\frac{2MG}{c^2r},~~a(r>R)=\frac{2MG}{c^2r}
\label{50}
\end{equation}                                 
for a source with rest energy $Mc^2=4\pi \int_0^Rdr^{\prime}r^{\prime 2}\rho(r^{\prime})$. 

With Eq. (\ref{46}) only differing from the second equation in Eq. (\ref{47}) by the replacement of $S_0^2/6$ by $-c^3/8\pi G$, any attempt to solve Eqs. (\ref{44}) -- (\ref{46}) via an analog of Eq. (\ref{49}) would immediately lead to us a Schwarzschild geometry with repulsive gravity. Since we wish to recover the exterior attractive gravity Eq. (\ref{50}) in the conformal case, we must thus not recover the analog of the interior Eq. (\ref{49}) as well, i.e. in the conformal case we need to find some different $r<R$ solution which will still match on to the exterior Eq. (\ref{50}). Noting now that Eq. (\ref{47}) involves $\nabla^2b$ but not $\nabla^2a$, we see that the $a(r)$ sector of the standard gravity Eq. (\ref{47}) does not involve any $\nabla^2(1/r)$ term with its associated $-4\pi\delta(r)/r^2$ singularity, with the $1/r$ term in $a(r)$ not needing any delta function source to support it. Consequently, returning now to the conformal case where there is equally no $\nabla^2a$ term in Eq. (\ref{46}), in the conformal case we can thus look for a solution in which $a(r)$ has a $1/r$ form for all $r$, both exterior and interior. For all $r$ we thus set
\begin{equation}          
a(r)=\frac{2d}{c^2r}-er
\label{51}
\end{equation}                                 
where $d$ and $e$ are as yet undetermined constants, and where the $er$ term that we also introduce will prove useful below. (With Eq. (\ref{44}) containing no $\nabla^4a$ term, the $er$ term is able to be present in $a(r)$ at all $r$ also, as it too will not generate any delta function singularity.) Quite remarkably, we find that for our candidate form for $a(r)$, the contribution of its $2d/c^2r$ term not only cancels identically in Eq. (\ref{46}), it also cancels identically in Eq. (\ref{44}) as well. In addition, the $er$ term also drops out of the left-hand side of Eq. (\ref{44}) identically, though it does generate terms of the form $eS_0^2/r$ elsewhere in these same equations. However, we shall show below that on solar system distance scales such $eS_0^2/r$ terms are totally negligible compared to the $\rho/c$ terms in Eqs. (\ref{44}) and (\ref{46}), with their neglect then bringing Eqs. (\ref{44}) and (\ref{46}) to the form 
\begin{equation}          
\nabla^4b=-\frac{3\rho}{4\alpha_gc},~~\nabla^2b=\frac{6\rho}{cS_0^2}.
\label{52}
\end{equation}                                 

With the generic forms of the solutions to these Poisson equations being given as in Eqs. (\ref{24}) and (\ref{49}), the exterior solution thus has to simultaneously be given as both 
\begin{equation}          
b(r>R)=\frac{3r}{8\alpha_{g}c}\int_0^R
dr^{\prime}r^{\prime 2}\rho(r^{\prime})
+\frac{1}{8\alpha_{g}cr}\int_0^R
dr^{\prime}r^{\prime 4}\rho(r^{\prime})
\label{53}
\end{equation}                                 
and 
\begin{equation}          
b(r>R)=-\frac{6}{S_0^2cr}\int_0^Rdr^{\prime}r^{\prime 2}\rho(r^{\prime}).
\label{54}
\end{equation}                                 
With the rest energy of the source being given by the positive
$Mc^2=4\pi \int_0^Rdr^{\prime}r^{\prime 2}\rho(r^{\prime})$, the solution of Eq. (\ref{54}) immediately assures us that  its associated Newtonian potential is indeed attractive. Compatibility of the Newtonian terms contained in the solutions of Eqs. (\ref{53}) and (\ref{54}) requires that the involved parameters be related via
\begin{equation}          
\frac{6Mc}{4\pi S_0^2}=-\frac{M_4c}{32 \pi \alpha_g},
\label{55}
\end{equation}                                 
(we use $M_4c^2=4\pi \int_0^Rdr^{\prime}r^{\prime 4}\rho(r^{\prime})$ to denote the fourth moment of the energy density), with the right-hand side of Eq. (\ref{55}) indeed being able to be positive for an appropriate choice of the sign of $\alpha_g$. Compatibility of Eqs. (\ref{53}) and (\ref{54}) additionally requires that the coefficient $\alpha_g$ be large enough to make the linear potential term $3rMc/32 \pi \alpha_g$ be negligible on solar system distance scales where we are looking for a linearized solution in the first place (i.e. as we had noted previously, since the fourth order theory leads to both Newtonian and linearly rising potentials, we can only  linearize in the region where $r$ is neither too small nor too big). Finally, on inserting the exterior solution of Eq. (\ref{53}) for $b(r)$ into Eq. (\ref{45}), the one equation still remaining, we see that in this solution the left-hand side of Eq. (\ref{45}) will vanish identically in the exterior region if we take the strengths of the linear potential terms in the exterior $a(r)$ and $b(r)$ to be equal and opposite (viz. $e=3Mc/32 \pi \alpha_g$), with the associated exterior vanishing of the right-hand side of Eq. (\ref{45}) (viz. none other than the vanishing of $G^r_{\phantom {r}r}$ in the exterior region) then fixing the parameter $2d/c^2$ in  the expression for $a(r)$ given in Eq. (\ref{51}) to be of the form $2d/c^2=6Mc/4\pi S_0^2$, just as required for the attractive exterior Schwarzschild solution. (The contributions of the linear potential term in $b(r)$ (viz. $b(r)=er$) to Eqs. (\ref{45}) and (\ref{46}) are also suppressed  in regions where $S_0^2e/r$ is negligible.) It is thus the freedom to assign both the magnitude and the sign of $\alpha_g$ which enables us to achieve compatibility of the solutions of Eqs. (\ref{53}) and (\ref{54}) on the solar system distance scale region of interest to us.

To show that there is at least one explicit choice for $\rho(r)$ for which the needed compatibility between the solutions of Eqs. (\ref{53}) and (\ref{54}) can be made manifest, guided by the highly singular structure of the source given in Eq. (\ref{27}) we set
\begin{equation}
\rho(r)=\frac{Mc^2}{4\pi} \frac{\delta(r)}{r^2}
+\frac{3vc}{2}\left[\nabla^2
-\frac{r^2}{12}\nabla^4\right]\left[\frac{\delta(r)}{r^2}\right],
\label{56}
\end{equation}                                 
to obtain as the two respective forms for $b(r>R)$
\begin{equation}          
b(r>R)=\frac{3Mcr}{32\pi \alpha_{g}}-\frac{3v}{4\alpha_{g}r}
\label{57}
\end{equation}                                 
and 
\begin{equation}          
b(r>R)=-\frac{3Mc}{2 \pi S_0^2r}.
\label{58}
\end{equation}                                 
In Eq. (\ref{58}), and central to the analysis here, we note that the carefully chosen $v$-dependent term in the source given in Eq. (\ref{56}) makes no contribution at all to the second moment integral needed in Eq. (\ref{54}), with the second order Poisson equation simply being totally insensitive to its possible existence. The identification of a local effective gravitational constant $G_{\rm LOC}$  according to
\begin{equation}          
\frac{2}{S_0^2}=+\frac{8\pi G_{\rm LOC}}{3c^3}
\label{59}
\end{equation}                                 
allows us to write Eq. (\ref{58}) and the $1/r$ part of Eq. (\ref{51}) as the attractive exterior metric
\begin{equation}          
b(r>R)=-\frac{2MG_{\rm LOC}}{c^2r},~~a(r>R)=\frac{2MG_{\rm LOC}}{c^2r}.
\label{60}
\end{equation}                                 
Compatibility of this solution with that of Eq. (\ref{57}) requires that the $v$ and $\alpha_g$ parameters be constrained according to
\begin{equation}          
\frac{3Mc}{2 \pi S_0^2}=\frac{2MG_{\rm LOC}}{c^2}=\frac{3 v}{4\alpha_g},
\label{61}
\end{equation}                                 
a relationship which can be satisfied by making an appropriate choice for the magnitude and sign of $v/\alpha_g$. (In a full non-perturbative treatment of Eqs. (\ref{5}) and (\ref{10}) the radial dependence of the energy density and pressure of the fluid would be self-consistently determined as part of the solution. Here we seek only to show that in the linearized case there is a choice for $\rho(r)$ for which one can obtain compatibility between the solutions of Eqs. (\ref{57}) and (\ref{58}).) 

As regards the numerical values of the various coefficients in our solution, for one solar mass of material we want the solution of Eq. (\ref{57}) to recover the $B(r>R)=1-2\beta/r+\gamma r$ solution whose fitting to galactic rotation curves described earlier requires that the coefficients take the numerical values $2\beta=2.96\times 10^5$ cm, $\gamma=5.42\times 10^{-41}$ cm$^{-1}$. For these particular values we find that if we take $G_{\rm LOC}$ to be equal to the standard Newtonian $G$, the parameters in Eqs. (\ref{57}) and (\ref{58}) are then given as $S_0^2=9.66\times 10^{37}$ gm sec$^{-1}$, $v= 1.29\times 10^{88}$ gm cm$^3$ sec$^{-1}$, $\alpha_g=3.29\times 10^{82}$ gm cm$^2$ sec$^{-1}$ (i.e. $\alpha_g$ has the dimension of action as required to make it dimensionless in natural units); and with these numbers we find that the $3Mcr/32\pi \alpha_{g}$ and $eS_0^2/r$ terms are indeed completely negligible on solar system distance scales where we are seeking a linearization of the theory (viz. $3Mcr/32\pi \alpha_{g} \ll 3Mc/2\pi S_0^2r$ and $eS_0^2/r \ll \rho/c \sim Mc/R^3$). Our analysis thus shows that even with a constant macroscopic scalar field in the region exterior to the sun, it is still possible for conformal gravity to be compatible with solar system Schwarzschild phenomenology, with it being central to our analysis that the source possess much deeper singularities than the sources that are ordinarily considered in the standard theory.

\section{Reconciliation of the calculations}

Since we have constructed an explicit solution for the linearized metric coefficients $b(r)$ and $a(r)$ using the standard coordinate system, it must be the case that on making a coordinate transform to isotropic coordinates the resulting metric coefficients $h(\rho)$ and $j(\rho)$ must provide a solution to the problem in the isotropic coordinate system. And yet inspection of Eq. (\ref{37}) would suggest otherwise. To reconcile this seeming contradiction we note that while the $b(r)$ and $a(r)$ coefficients respectively transform into $h(\rho)$ and $j(\rho)$, because the transformation between the full non-linearized metrics of Eqs. (\ref{30}) and (\ref{39}) is of the form
\begin{equation}          
r=\rho J^{1/2}(\rho),~~A^{-1/2}(r)= 1+\frac{\rho J^{\prime}(\rho)}{2J(\rho)},~~B(r)=H(\rho),
\label{62}
\end{equation}                                 
in this  transformation $A(r)$ is related not to $J(\rho)$ but to a derivative of it instead. Thus,  what had initially been a first derivative function of $a(r)$ in Eq. (\ref{46}) becomes a second derivative function of $j(\rho)$ in Eq. (\ref{37}). Specifically, with the linearized transform itself being of the form
\begin{equation}          
r \rightarrow \rho +\frac{\rho j(\rho)}{2},~~-a(r) \rightarrow \rho j^{\prime}(\rho),~~b(r)\rightarrow h(\rho),
\label{63}
\end{equation}                                 
the $-2a/r^2-2a^{\prime}/r$ combination which appears in the expression for the Ricci scalar given in Eq. (\ref{46}) thus transforms into the $2\nabla^2j(\rho)$ term which appears in the expression for the Ricci scalar given in Eq. (\ref{37}), with Eq. (\ref{63}) explicitly leading to 
\begin{equation}          
-\frac{2a^{\prime}}{r}-\frac{2a}{r^2} \rightarrow 2\left[j^{\prime\prime}+\frac{2j^{\prime}}{\rho}\right] =2\nabla^2j
\label{64}
\end{equation}                                 
in lowest order. Now the $1/r$ part of our chosen standard coordinate system solution for $a(r)$ as given by $a(r)=2d/c^2r$ has the property that it causes the quantity $-2a/r^2-2a^{\prime}/r$ to vanish identically. Hence it must thus be the case that its transform causes the $\nabla^2j(\rho)$ term to vanish identically too. However, on transforming $a(r)=2d/c^2r$ we find that it transforms into $j(\rho)=2d/c^2\rho$, with insertion of this form for $j(\rho)$ into $\nabla^2j$ yielding the non-vanishing $\nabla^2(2d/c^2\rho)=-(8\pi d/c^2)\delta(\rho)/\rho^2$. However, this contradiction is only an apparent one since the quantity $\nabla^2(2d/c^2\rho)$ only becomes singular at $\rho=0$, a region where $j(\rho)$ as given by $2d/c^2\rho$ becomes too large for a linearized approximation to hold in the first place. The apparent non-vanishing of $\nabla^2j$ is thus a spurious artifact of the linearization in the isotropic coordinate system, with $j(\rho)$ being able to be equal to $2d/c^2\rho$ for all not too small $\rho$, and with the $\nabla^2j(\rho)$ term which appears in the Ricci scalar term in the left-hand side Eq. (\ref{37}) then not coupling to the $\rho(\rho)$ source which appears on its right-hand side at all. Thus given this pecularity  of the isotropic coordinate system, we must solve Eq. (\ref{37}) by setting the $\nabla^2j$ term equal to zero in it (while likewise setting the $\nabla^4j$ term equal to zero in Eq. (\ref{35})),  to then reduce Eqs. (\ref{37}) and (\ref{35}) to
\begin{equation}             
\frac{S_0^2}{6}\nabla^2h=\frac{\rho(\rho)}{c},~~-\frac{4\alpha_g}{3}\nabla^4h=\frac{ \rho(\rho)}{c},
\label{65}
\end{equation}                                 
to thereby yield attractive gravity after all; with the full solution to the isotropic coordinate system Eqs. (\ref{35}) - (\ref{37}) in the region exterior to a source of radius $\hat{\rho}$ then being given by
\begin{equation}             
h(\rho>\hat{\rho})=\frac{3Mc\rho}{32\pi \alpha_{g}}-\frac{3v}{4\alpha_{g}\rho},~~h(\rho>\hat{\rho})=-\frac{3Mc}{2 \pi S_0^2\rho},~~j(\rho>\hat{\rho})=\frac{3Mc}{2 \pi S_0^2\rho}+\frac{3Mc\rho}{32\pi \alpha_{g}},
\label{66}
\end{equation}                                 
in complete analog to the standard coordinate system solution given previously.

Pedagogically, it is instructive to explore this peculiarity of isotropic coordinate system linearization a little further. If we return to the full non-linearized theory, we can write closed form expressions for the $G^t_{\phantom{t}t}$ component of the Einstein tensor in both the standard (${\rm STA}$) and the isotropic (${\rm ISO}$) coordinate systems, to respectively obtain 
\begin{equation}             
G^t_{\phantom{t}t}({\rm STA})=\frac{A^{\prime}}{rA^2}+\frac{1}{r^2}-\frac{1}{r^2A},
\label{67}
\end{equation}                                 
and 
\begin{equation}             
G^t_{\phantom{t}t}({\rm ISO})=-\frac{J^{\prime\prime}}{J^2}-\frac{2J^{\prime}}{\rho J^2} +\frac{3J^{\prime 2}}{4J^3},
\label{68}
\end{equation}                                 
with these two expression being related by the exact coordinate transform given in Eq. (\ref{62}), a transformation which converts the first derivative function $G^t_{\phantom{t}t}({\rm STA})$ into the second derivative function $G^t_{\phantom{t}t}({\rm ISO})$.
 (In Eq. (\ref{67}) the primes denote derivatives with respect to $r$ and in Eq. (\ref{68}) they denote derivatives with respect to $\rho$.) For our purposes here it is more convenient to replace $J(\rho)$ by the  form 
\begin{equation}             
J(\rho)=\left[1+K(\rho)\right]^4,
\label{69}
\end{equation}                                 
as the isotropic coordinate  coordinate $G^t_{\phantom{t}t}({\rm ISO})$ then gets rewritten as
\begin{equation}             
G^t_{\phantom{t}t}({\rm ISO})=-\frac{4}{(1+K)^5}\left[K^{\prime\prime} +\frac{2K^{\prime}}{\rho}\right].
\label{70}
\end{equation}                                 

 If we now consider an Einstein theory with no source at all, i.e. if we simply set $G^t_{\phantom{t}t}$ equal to zero everywhere, we then find exact solutions to the equation $G^t_{\phantom{t}t}=0$ in the two coordinate systems to be of the form
\begin{equation}             
A(r)=\left(1-\frac{2C}{r}\right)^{-1},
\label{71}
\end{equation}                                 
\begin{equation}
J(\rho)=\left(1+\frac{C}{2\rho}\right)^4~~,~~K(\rho)=\frac{C}{2\rho},            
\label{72}
\end{equation}                                 
where $C$ is an unspecified constant and where the two solutions are characterized by the same parameter $C$ as they are related by the exact coordinate transform given in Eq. (\ref{62}).  That this form for $K(\rho)$ really is a solution to $G^t_{\phantom{t}t}({\rm ISO})=0$ is due to the fact that while the action of the $K^{\prime\prime} +2K^{\prime}/\rho$ term on $C/2\rho$ gives the non-zero $-2C\pi \delta(\rho)/\rho^2$, the prefactor $1/(1+K)^5$ vanishes as $\rho^5$ near $\rho=0$ to then cancel the delta function. However, suppose we now linearize the theory around flat spacetime to obtain
\begin{equation}             
G^t_{\phantom{t}t}({\rm STA;LIN})=\frac{A^{\prime}}{r}+\frac{(A-1)}{r^2},
\label{73}
\end{equation}                                 
\begin{equation}             
G^t_{\phantom{t}t}({\rm ISO;LIN)}=-4\left[K^{\prime\prime} +\frac{2K^{\prime}}{\rho}\right].
\label{74}
\end{equation}                                 
Setting $G^t_{\phantom{t}t}({\rm STA;LIN})$ equal to zero would still possess a solution for all $r$, viz. 
\begin{equation}             
A(r)=1+\frac{2C}{r},
\label{75}
\end{equation}                                 
but the function into which it would transform via a coordinate transformation between the two coordinates systems, viz.
\begin{equation}             
J(\rho)=1+\frac{2C}{\rho},~~K(\rho)=\frac{C}{2\rho}, 
\label{76}
\end{equation}                                 
would not satisfy $K^{\prime\prime} +2K^{\prime}/\rho=0$ at $\rho=0$ because of the delta function term that it would generate in it. Consequently, even while the solution of Eq. (\ref{75}) is a solution to $G^t_{\phantom{t}t}({\rm STA;LIN})=0$ for all $r$, the function into which it transforms is not a solution to $G^t_{\phantom{t}t}({\rm ISO;LIN})=0$ for all $\rho$, with the reason for this being that in its linearization $G^t_{\phantom{t}t}({\rm ISO})$ loses its crucial $1/(1+K)^5$ prefactor. Thus, with the coordinate transformation of Eq. (\ref{62}) relating derivatives of different orders,  linearization in an isotropic coordinate system is not as straightforward as linearization in a standard coordinate system, with it being the linearized standard coordinate system which is best suited to accommodate singularities.

\section{On the physical significance of the scalar field}

While we have now achieved our primary purpose of showing that the standard Schwarzschild phenomenology can in fact be recovered in conformal gravity in the presence of a macrosopic scalar field, it is nonetheless instructive to discuss the physical significance of the scalar field and its relation first to dynamical mass generation and then to dynamical localization of particles in conformally invariant theories. It is also instructive to explore the circumstances under which the $-S^2R^{\mu}_{\phantom{\mu}\mu}/12$ term in the action of Eq. (\ref{2}) could in fact lead to repulsive gravity. While we have, for simplicity, used a fundamental scalar field for the analysis presented in this paper, in the dynamical case the scalar field would be the vacuum expectation value of a fermion composite operator such as the fermion mass generating bilinear $\bar{\psi}(x)\psi(x)$. And whether or not the expectation value of any such composite is to be constant or spatially dependent will depend on the state in which the expectation value is to be taken. Moreover, with the discussion of dynamical symmetry breaking by fermion composites having been formulated primarily in flat spacetime, we will need to adapt the formulation to curved spacetime.  

With regard first to the discussion of dynamical symmetry breaking in flat spacetime itself,  following the pioneering work of Nambu and Jona-Lasinio \cite{Nambu1961} based on analogy with the BCS theory of superconductivity, we recall that in order to find the ground state of an interacting fermionic system one looks at states in which all the negative energy fermionic states are occupied (for illustrative purposes in the following we shall denote the number of such occupied states as $N$), with all the positive energy fermionic states being unoccupied. Two candidate ground states are suggested: the normal one $|N\rangle$ in which the fermion is massless and the bilinear expectation value $\langle N|\bar{\psi}(x)\psi(x)|N\rangle$ is zero, and the self-consistently determined spontaneously broken superconducting  type state $|S\rangle$ in which the fermion is massive and the bilinear expectation value $\langle S|\bar{\psi}(x)\psi(x)|S\rangle$ is non-zero. Dynamical symmetry breaking then occurs in a  theory with no fundamental scale whenever the dynamics is such that the state $|S\rangle$ has lower energy than the state $|N\rangle$, with the non-vanishing of the expectation value $\langle S|\bar{\psi}(x)\psi(x)|S\rangle$ then introducing a scale dynamically. If such a state $|S\rangle$ is to be the vacuum of the theory, then with the ground state needing to be translation invariant, the expectation value $\langle S|\bar{\psi}(x)\psi(x)|S\rangle$ would then be independent of the coordinate $x_{\mu}$ and thus be constant throughout the spacetime. As such, the lowering of the energy from that in the state $|N\rangle$ (viz. $\langle N|H|N\rangle$ where $H$ is the Hamiltonan) to that in the state $|S\rangle$ (viz. $\langle S|H|S\rangle$) would yield a cosmological constant term which while unimportant in flat spacetime (where only energy differences are observable) becomes central once gravity is introduced since gravity couples to energy itself.

With respect to such a state $|S\rangle$ one can introduce creation and annihilation operators for the fermion (generically $a^{\dagger}$ and $a$), with the annihilation operators then annihilating the state $|S\rangle$ according to $a|S\rangle =0$. Given such a Fock space it is very tempting to identify the 
$N+1$ particle state $a^{\dagger}|S\rangle$ as the lowest lying positive energy fermionic mode, a mode which propagates as a translation invariant plane wave above a Dirac sea filled with $N$ negative energy plane wave modes. However, such an identification is not necessarily correct since the state $|S\rangle$ was determined to be the self-consistent $N$ particle state, with the state  $a^{\dagger}|S\rangle$ not automatically being a self-consistent $N+1$ particle eigenstate of the Hamiltonian. Rather, the $N+1$ particle state needs to be calculated self-consistently all over again, with it being found in certain dynamical cases (the double-well potential fundamental scalar field case \cite{Dashen1974,Bardeen1975} and a Nambu-Jona-Lasinio  type fermion composite model \cite{Dashen1975}) that the $N+1$ particle state could lower its energy with respect to the state $a^{\dagger}|S\rangle$ by becoming spatially dependent, with this spatially dependent state then being the self-consistent $N+1$ particle eigenstate of the Hamiltonian. Specifically, one constructs a spatially dependent $N$ particle coherent state $|C\rangle$ in which the expectation value $\langle C|\bar{\psi}(x)\psi(x)|C\rangle$ is spatially dependent ($|C\rangle$ is constructed from $|S\rangle$ via a spatially dependent Bogolubov transform on it), one defines creation and annihilation operators (generically $b^{\dagger}$ and $b$) with respect to $|C\rangle$ so that $b|C\rangle=0$,  and then constructs the $N+1$ particle state $b^{\dagger}|C\rangle$. Such an $N+1$ particle state is not built out of plane waves. Rather, all the $N$ negative energy modes are distorted in the vicinity of some center of localization, with the lowest positive energy state then being bound to these distorted negative energy states with a wave function which is highly localized and which falls off very fast as we go away from the center of localization. Moreover, the state $|C\rangle$ itself need not be an $N$ particle eigenstate at all. Rather, it is the $N+1$ particle $b^{\dagger}|C\rangle$ which is the eigenstate, being stabilized through a cooperative effect between the positive energy fermion and the filled negative energy sea in which the localizing of the positive energy fermion distorts the wave functions of the fermions in the negative energy sea causing them to form a localized potential in which the positive energy fermion is then bound.  With the spatially independent state $|S\rangle$ being the ground state of the system rather than the spatially dependent state $|C\rangle$, there is no violation of translation invariance, with it thus being possible to produce localized excited states in a conformal invariant theory without needing to break translation invariance. 

While the negative energy modes are distorted in the vicinity of the center of localization in such coherent states, far from the center of localization their wave functions revert back to plane waves. Consequently, asymptotically far from the center of localization the spatially dependent expectation value $\langle C|\bar{\psi}(x)\psi(x)|C\rangle$ will approach the constant value associated with the expectation value $\langle S|\bar{\psi}(x)\psi(x)|S\rangle$. If we thus represent  $\langle C|\bar{\psi}(x)\psi(x)|C\rangle$ by a spatially dependent order parameter $S(x)$ and represent  $\langle S|\bar{\psi}(x)\psi(x)|S\rangle$ by a spatially independent order parameter $S_0$, we see that $S(x)$ will approach $S_0$ asymptotically far from the center of localization. Moreover, not only will such an $S(x)$ approach $S_0$ asymptotically, in the flat spacetime dynamical symmetry breaking models studied in \cite{Dashen1975,Mannheim1976} the approach is exponentially fast (c.f. the dynamical kink type states of \cite{Dashen1975} where $S(x)\sim S_0{\rm tanh}(Mx)$, with $S(x)$ behaving asymptotically as $S_0(1-O(e^{-2Mx}))$ where $M$ is the dynamically induced fermion mass). Thus while $S(x)$ itself might be large, its derivatives would fall off very fast. This then is the localization mechanism for dynamical symmetry breaking in flat spacetime. 

In extending these ideas to curved space, we need to interface them with early universe cosmology since mass generating phase transitions occur as the universe cools down, with the fermion masses and the fermion condensate $S(x)$ being thought to be associated with the electroweak symmetry breaking phase transition which occurs at a temperature $T_{\rm EW}$ of the order of $10^{15}$ $^\circ K$ or so. However, even above such temperatures there are already dynamical mass scales present such as for instance those associated with the expansion radius and scalar curvature of the universe. To generate scales such as these there should thus also be a much higher temperature early universe phase transition at some generic grand-unified type scale temperature  $T_{\rm GUT}$ of the order of $10^{28}$ $^\circ K$ or so. Since this much higher temperature phase transition does not generate fermion masses, its associated condensate should not Yukawa couple to fermions, with a natural candidate for the needed condensate thus being the expectation value of a fermion quadrilinear $\bar{\psi}(x)\psi(x)\bar{\psi}(x)\psi(x)$ in some `ur' state $|U\rangle$ with expectation value $U(x)=\langle U|\bar{\psi}(x)\psi(x)\bar{\psi}(x)\psi(x)|U\rangle$. The state $|U\rangle$ thus has the property that at temperatures above $T_{\rm EW}$ the expectation value $\langle U|\bar{\psi}(x)\psi(x)|U\rangle$ of the fermion bilinear is zero. While these general remarks hold in both standard cosmology and conformal gravity cosmology, in the conformal case, just like $S(x)$, $U(x)$ needs to be conformally coupled  to gravity with the full energy-momentum tensor needing to be traceless. In the presence of the two fields $U(x)$ and $S(x)$ we thus replace the conformal matter action $I_M$ of Eq. (\ref{2}) by 
\begin{eqnarray}
I_M&=&-\int d^4x(-g)^{1/2}\bigg{[}\frac{1}{2}S^{;\mu}
S_{;\mu}-\frac{1}{12}S^2R^\mu_{\phantom         
{\mu}\mu}
+\lambda S^4
+i\bar{\psi}\gamma^{\mu}(x)[\partial_\mu+\Gamma_\mu(x)]             
\psi -hS\bar{\psi}\psi
\nonumber \\
&&+\frac{1}{2}U^{;\mu}
U_{;\mu}-\frac{1}{12}U^2R^\mu_{\phantom         
{\mu}\mu}
+\tau U^4\bigg{]},
\label{77}
\end{eqnarray}                                 
with the full $T^{\mu\nu}$ of Eqs. (\ref{7}) and (\ref{8}) being generalized to 
\begin{eqnarray}
T^{\mu \nu} &=& \frac{1}{c}\left[(\rho+p)U^{\mu}U^{\nu}+pg^{\mu\nu}\right]
+\frac{2}{3}S^{;\mu} S^{;\nu} 
-\frac{1}{6}g^{\mu\nu}S^{;\alpha} S_{;\alpha}
-\frac{1}{3}SS^{;\mu;\nu}
+\frac{1}{3}g^{\mu\nu}SS^{;\alpha}_{\phantom{;\alpha};\alpha}  
\nonumber \\             
&&                          
-\frac{1}{6}S^2\left(R^{\mu\nu}
-\frac{1}{2}g^{\mu\nu}R^\alpha_{\phantom{\alpha}\alpha}\right)         
-g^{\mu\nu}\lambda S^4 
+\frac{2}{3}U^{;\mu} U^{;\nu} 
-\frac{1}{6}g^{\mu\nu}U^{;\alpha} U_{;\alpha}
-\frac{1}{3}UU^{;\mu;\nu}
\nonumber \\
&&+\frac{1}{3}g^{\mu\nu}UU^{;\alpha}_{\phantom{;\alpha};\alpha} -\frac{1}{6}U^2\left(R^{\mu\nu}
-\frac{1}{2}g^{\mu\nu}R^\alpha_{\phantom{\alpha}\alpha}\right)         
-g^{\mu\nu}\tau U^4, 
\label{78}
\end{eqnarray}                                 
and with Eq. (\ref{10}) being extended to the two scalar field equations
\begin{equation}
\frac{1}{6}S^2R^\mu_{\phantom{\mu}\mu}
=\frac{1}{c}\left[\rho-3p\right]+4\lambda S^4 -SS^{;\mu}_{\phantom{\mu};\mu},
\label{79}
\end{equation}
\begin{equation}
\frac{1}{6}U^2R^\mu_{\phantom{\mu}\mu}
=4\tau U^4 -UU^{;\mu}_{\phantom{\mu};\mu}.
\label{80}
\end{equation}                                 

In the explicit application of conformal gravity to cosmology, we need to take into consideration both the background cosmology and the fluctuations around it. The background cosmology is homogeneous and highly symmetric being  either Robertson-Walker or de Sitter (for simplicity in the following we shall take the background to be de Sitter), while the fluctuations around it are inhomogeneous and far less symmetric. (While  conformal gravity has no apparent need for inflation since its Robertson-Walker phase is free of both the flatness and the horizon problems \cite{Mannheim2006}, a de Sitter cosmology is nonetheless still an allowed solution in the conformal case.) The background geometry can thus be modeled by constant values for the $U(x)$ and $S(x)$ fields, a background in which the fermions propagate as plane waves,  while the inhomogeneities can be described by fermions localized in coherent states in which the positive energy fermion is localized into some finite region $r<R$ and in which the negative energy fermion wave functions only approach plane waves asymptotically far from the center of localization. However, since both the fermion bilinear and quadrilinear condensates are built out of one and the same set of fermions, the distortion of the fermion wave functions in the vicinity of the center of localization will cause both $U(x)$ and $S(x)$ to depart locally from their asymptotic background $U_0$ and $S_0$ values. Since in principle these two departures are not the same, and since the two fields transform independently under a conformal transformation, any local conformal transformation which might bring one of them to a constant throughout the spacetime would not simultaneously bring the other one to a constant form as well. We thus cannot take both $U(x)$ and $S(x)$ to be constant, and so in the following we shall in fact allow both of them to acquire a spatial dependence once an inhomogeneity is introduced, and shall use these spatial dependences to explicitly monitor the modifications to the geometry caused by the introduction of the inhomogeneity. With the de Sitter background geometry actually being writable in the static coordinate form given in Eq. (\ref{20}), for the case of a single mass source localized in a de Sitter background the equations of motion of Eqs. (\ref{22}) and (\ref{29}) thus generalize to
\begin{equation}                                                                              
\nabla^4B=f(r) = \frac{1}{4\alpha_g }\left[-\frac{3(\rho+p)}{cB}
+SS^{\prime \prime}-2S^{\prime 2} +UU^{\prime \prime}-2U^{\prime 2}\right],                  
\label{81}
\end{equation}                                                  
\begin{equation}
\frac{1}{6}S^2\left[B^{\prime\prime}+\frac{4B^{\prime}}{r}+\frac{2B}{r^2}-\frac{2}{r^2}\right] =\frac{ (\rho -3p)}{c}+4\lambda S^4 -S\left[BS^{\prime\prime}+\frac{2BS^{\prime}}{r}+B^{\prime}S^{\prime}\right],
\label{82}
\end{equation}
\begin{equation}
\frac{1}{6}U^2\left[B^{\prime\prime}+\frac{4B^{\prime}}{r}+\frac{2B}{r^2}-\frac{2}{r^2}\right] =4\tau U^4 -U\left[BU^{\prime\prime}+\frac{2BU^{\prime}}{r}+B^{\prime}U^{\prime}\right],
\label{83}
\end{equation}
to thereby define the model. As we see, the great utility of allowing both $U(r)$ and $S(r)$ to be spatially dependent is that we can take full advantage of the conformal structure of the metric of Eq. (\ref{19}) to obtain equations which are linear in the metric coefficient $B(r)$, regardless in fact of the magnitude of $B(r)$, and thus regardless of whether or not the geometry might be close to flat. What prevents finding analytic solutions to these equations is only that they are non-linear in the scalar fields.

With a background de Sitter (or Robertson-Walker) geometry being conformal to flat, in such a geometry the Weyl tensor vanishes, and thus also the tensor $W^{\mu\nu}$ introduced in Eq. (\ref{6}). In a de Sitter geometry then, if we set take both $U(r)$ and $S(r)$ to be constant, we find that Eqs. (\ref{81}) - (\ref{83}) admit of the exact de Sitter geometry solution
\begin{equation}
B(r)=1-kr^2,~~\rho_0+p_0=0,~~-2kS_0^2=4\lambda S_0^4 +\frac{(\rho_0-3p_0)}{c},~~-2kU_0^2=4\tau U_0^4,
\label{84}
\end{equation}
where the spatially independent $\rho_0$ and $p_0$ denote the background values of $\rho$ and 
$p$. (The de Sitter solution is an example of a solution to Eqs. (\ref{81}) - (\ref{83}) in which the geometry is far from flat even though the equations of motion are linear in $B(r)$.) As such, the solution of Eq. (\ref{84}) not only defines the background, in it we see that the explicit removal of the conformal factor $p^2(r)/r^2$ in going from the metric of Eq. (\ref{19}) to that of Eq. (\ref{20}) then gives us a form of the metric which precisely is de Sitter, viz.
\begin{equation}
ds^2= -(1-kr^2)c^2 dt^2+ {dr^2\over (1-kr^2)} + r^2d\Omega_2,
\label{85}
\end{equation}
in the solution. Use of the metric of Eq. (\ref{20}) thus precisely dovetails with the use of constant background scalar fields. And, moreover, as we see, one should not drop the $4\lambda S_0^4$ term as had been done in Eq. (\ref{11}) since this term supports the $-(S_0^2/6)(R^{\mu\nu}
-\frac{1}{2}g^{\mu\nu}R^\alpha_{\phantom{\alpha}\alpha})$ term that was retained. Finally, we note that  since the Schwarzschild-de Sitter metric can also be written in the form of the metric given in Eq. (\ref{20}) (with $B(r)=1-2\beta/r-kr^2$), Eqs. (\ref{81}) - (\ref{83}) are thus ideal for monitoring inhomogeneities in a de Sitter background. 

However, before doing this, we note that since $W^{\mu\nu}$ does vanish in a de Sitter or Robertson-Walker background, then, according to Eq. (\ref{5}),  in such a background the full $T^{\mu\nu}$ of Eq. (\ref{78}) must vanish also, with its vanishing then yielding the equation of motion. 
\begin{equation}
\frac{1}{6}[U_0^2+S_0^2]\left(R^{\mu\nu}
-\frac{1}{2}g^{\mu\nu}R^\alpha_{\phantom{\alpha}\alpha}\right) = \frac{1}{c}\left[(\rho+p)U^{\mu}U^{\nu}+pg^{\mu\nu}\right] -g^{\mu\nu}\tau U_0^4 -g^{\mu\nu}\lambda S_0^4
\label{86}
\end{equation}
when the scalar fields assume their background values. As such, we recognize Eq. (\ref{86}) to be in the form of none other than the standard Friedmann cosmological evolution equation, save only that the standard attractive Newton constant has been replaced \cite{Mannheim1992} by a global effective repulsive gravitational constant $G_{\rm GLOB}$ of the form 
\begin{equation}          
\frac{8\pi G_{\rm GLOB}}{3c^3}=-\frac{2}{(U_0^2+S_0^2)}.
\label{87}
\end{equation}                                 
In conformal gravity then wrong sign terms such as $-U^2R^{\mu}_{\phantom{\mu}\mu}/12$ and $-S^2R^{\mu}_{\phantom{\mu}\mu}/12$ really do lead to a gravity which is repulsive, but on global rather than local scales. Moreover, this was found to be an advantage as it leads \cite{Mannheim2006} to a cosmology with no flatness problem, no horizon problem, no universe age problem, to a cosmology which is naturally accelerating without fine-tuning, and to a solution to the cosmological constant problem; with it not being the cosmological constant which is quenched down from the large value that would be associated with a large $U_0$ and $S_0$, but rather it is the amount (viz. $G_{\rm GLOB}$) by which it gravitates which is quenched instead, as the very same mechanism which makes $U_0$ and $S_0$ large simultaneously makes $G_{\rm GLOB}$ small,  with the conformal theory then being able to naturally fit \cite{Mannheim2006} the accelerating universe supernovae data without any fine-tuning at all despite using a cosmological constant which is as large as particle physics suggests. 

With mass sources only being able to localize in the presence of inhomogeneities, the local gravity produced by their presence will be associated with departures of the scalar fields from their constant background values and with departures of the Weyl tensor from its zero background value. In such a situation local gravitational departures from the background geometry will be controlled by the sign of $\alpha_g$, with there then being an induced local $G_{\rm LOC}$ whose sign is completely decoupled from that of the global $G_{\rm GLOB}$, with the $-U^2R^{\mu}_{\phantom{\mu}\mu}/12$ and $-S^2R^{\mu}_{\phantom{\mu}\mu}/12$ terms thus not leading to repulsive local gravity. To see exactly how it works in the conformal theory we thus need to look at the non-leading terms in $U(r)$ and $S(r)$, terms which will explicitly be spatially dependent. 

While we had noted above that in the flat space case departures from constant scalar field are exponential, an explicit study of Einstein gravity coupled to a scalar field has found \cite{Mannheim1991} that in the gravitational case such departures are only power behaved. Specifically it was found that an Einstein gravity theory coupled to the kinetic energy of a minimally coupled massless scalar field, viz. one based on the action
\begin{equation}
I=-\int d^4x (-g)^{1/2}\left[\frac{\kappa}{2}R^{\mu}_{\phantom{\mu}\mu}+\frac{1}{2}\partial^{\mu}S\partial_{\mu}S\right]
\label{88}
\end{equation}
and equations of motion
\begin{equation}          
-\kappa\left[R^{\mu\nu}
-\frac{1}{2}g^{\mu\nu}R^\alpha_{\phantom{\alpha}\alpha}\right] = \partial^{\mu}S\partial^{\nu}S -\frac{1}{2}g^{\mu\nu}\partial^{\alpha}S\partial_{\alpha}S,~~(-g)^{-1/2}\partial_{\mu}\left[(-g)^{1/2}\partial^{\mu}S\right]=0,
\label{89}
\end{equation}                                 
admitted the exact exterior isotropic coordinate system solution \cite{Buchdahl1959,Mannheim1991}
\begin{equation}          
S(\rho)=\frac{C}{2\hat{\rho}}{\rm ln}\left(\frac{\rho -\hat{\rho}}{\rho +\hat{\rho}}\right) +S_0,~~H(\rho)=\left(\frac{\rho -\hat{\rho}}{\rho +\hat{\rho}}\right)^{d/2\hat{\rho}},~~J(\rho)=\frac{(\rho^2 -\hat{\rho}^2)^2}{\rho^4}\left(\frac{\rho -\hat{\rho}}{\rho +\hat{\rho}}\right)^{-d/2\hat{\rho}}
\label{90}
\end{equation}                                 
where the parameter $d$ is given by
\begin{equation}          
d=\left(16\hat{\rho}^2-\frac{2C^2}{\kappa}\right)^{1/2}
\label{91}
\end{equation}                                 
and where $C$, $\hat{\rho}$ and $S_0$ are appropriate integration constants. As such this solution itself generalizes the point source solution to the same theory found even earlier by  Yilmaz \cite{Yilmaz1958} in which the integration constant $\hat{\rho}$ was zero, with the non-vanishing of $\hat{\rho}$ in Eq. (\ref{90}) enabling the solution to describe an extended source with a radius of order $\hat{\rho}$ rather than a point source. On setting the parameter $d$ equal to $2MG/c^2$, and converting the solution to the standard coordinate system of Eq. (\ref{39}), at large values of the radial coordinate $r$ the metric coefficients are found to behave as
\begin{eqnarray}          
B(r) &&\rightarrow 1 -\frac {2MG}{c^2r}+\frac{4MG\hat{\rho}^2}{3c^2r^3}-\frac{M^3G^3}{3c^6r^3},
\nonumber\\
\frac{1}{A(r)} &&\rightarrow 1 -\frac {2MG}{c^2r}+\frac{4\hat{\rho}^2}{r^2} -\frac{M^2G^2}{c^4r^2} +\frac{4MG\hat{\rho}^2}{c^2r^3}-\frac{M^3G^3}{c^6r^3},
\label{92}
\end{eqnarray}                                 
to thus yield power law corrections to the Schwarzschild metric. Moreover, it was further shown in \cite{Mannheim1991} that if the action of Eq. (\ref{88}) were to be augmented by a double-well Higgs potential of the form $V(S)=\lambda S^4-\mu^2S^2/2$ with its familiar minimum at $S_0=\mu/(4\lambda)^{1/2}$, the corrections to the now Schwarzschild-de Sitter metric would continue to be power behaved. With such corrections being only power behaved rather than exponential, it was further noted in \cite{Mannheim1991} that if a source such as the sun actually did get its mass dynamically via a macroscopic Higgs scalar field, the Higgs mechanism could actually be tested gravitationally by  monitoring these power law corrections to the Schwarzschild geometry, as the corrections associated with the expansion of Eq. (\ref{92}) could potentially be quite significant if the parameter $\hat{\rho}$ is substantially larger than the Schwarzschild radius of the sun, something that would indeed be the case if $\hat{\rho}$ is of order the radius of the sun.

\section{Power-behaved solution to the full conformal theory}

Given the above remarks, we shall thus seek power-behaved corrections to the conformal gravity background solution of Eq. (\ref{84}), and since we are in the fourth order theory we shall need to incorporate the linear potential term of the solution of Eq. (\ref{25}) as well. We thus look to see whether Eqs. (\ref{81}) - (\ref{83}) can support  a power-behaved exterior solution of the form
\begin{eqnarray}          
B(r>R)&=&-kr^2+\gamma r+w -\frac{2\beta}{r}+\frac{\delta}{r^2}+\ldots,
\nonumber \\
S(r>R)&=&S_0+\frac{S_1}{r} +\frac{S_2}{r^2} +\frac{S_3}{r^3}+\frac{S_4}{r^4}+\frac{S_5}{r^5}+\ldots,
\nonumber \\
U(r>R)&=&U_0+\frac{U_1}{r} +\frac{U_2}{r^2} +\frac{U_3}{r^3}+\frac{U_4}{r^4}+\frac{U_5}{r^5}+\ldots
\label{93}
\end{eqnarray}                                 
in the event that there is some localized source within the  $r<R$ region, a source whose very presence will also lead to exterior region corrections to the background global $\rho(r)$ and $p(r)$ (by distorting the plane wave modes of the negative energy fermion sea). As we shall see, these corrections to the perfect fluid energy density and pressure will be power behaved too, with the full energy density and pressure behaving as 
\begin{eqnarray}          
\rho(r)&=&\theta(R-r)\rho_{\rm LOC}(r)+\theta(r-R)\rho_{\rm GLOB}(r)
\nonumber \\
&=&\theta(R-r)\rho_{\rm LOC}(r)+ \theta(r-R)\left[\rho_0+\frac{\rho_1}{r} +\frac{\rho_2}{r^2} +\frac{\rho_3}{r^3}+\frac{\rho_4}{r^4}+\frac{\rho_5}{r^5}+\ldots\right],
\nonumber \\
p(r)&=&\theta(R-r)p_{\rm LOC}(r)+\theta(r-R)p_{\rm GLOB}(r)
\nonumber \\
&=&\theta(R-r)p_{\rm LOC}(r)+\theta(r-R)\left[p_0+\frac{p_1}{r} +\frac{p_2}{r^2} +\frac{p_3}{r^3}+\frac{p_4}{r^4}+\frac{p_5}{r^5}+\ldots\right],
\label{94}
\end{eqnarray}                                 
where $\rho_{\rm LOC}(r)$ and $p_{\rm LOC}(r)$ are the localized $r<R$ contributions to the energy density and pressure due to the presence of the localized positive energy fermion.

To see how the theory is able to support power-behaved corrections we need to insert the expansions of Eqs. (\ref{93}) and (\ref{94}) into the relevant quantities which appear in Eqs. (\ref{81}) - (\ref{83}), to obtain first for Eq. (\ref{82}) (and analogously for Eq. (\ref{83}))
\begin{eqnarray}
&&\frac{1}{6}S^2\bigg{[}B^{\prime\prime}+\frac{4B^{\prime}}{r}+\frac{2B}{r^2}-\frac{2}{r^2}\bigg{]} +S\bigg{[}BS^{\prime\prime}+\frac{2BS^{\prime}}{r}+B^{\prime}S^{\prime}\bigg{]}
\nonumber \\
&&=-2kS_0^2+\frac{1}{r}\bigg{[}-2kS_0S_1+\gamma S_0^2\bigg{]}
+\frac{1}{r^2}\bigg{[}-2kS_0S_2+\gamma S_0S_1+\frac{(w-1)}{3}S_0^2\bigg{]}
\nonumber \\
&&~~+\frac{1}{r^3}\bigg{[}-4kS_0S_3+2\gamma S_0S_2+\frac{2(w-1)}{3}S_0S_1\bigg{]}
+\frac{1}{r^4}\bigg{[}-8kS_0S_4-2kS_1S_3
\nonumber \\
&&~~+5\gamma S_0S_3+\gamma S_1S_2+\frac{2(w-1)}{3}S_0S_2+\frac{(w-1)}{3}S_1^2
+2wS_0S_2-2\beta S_0S_1\bigg{]}
\nonumber \\
&&~~+\frac{1}{r^5}\bigg{[}-14kS_0S_5-6kS_1S_4-2kS_2S_3+10\gamma S_0S_4+4\gamma S_1S_3  +\gamma S_2^2+\frac{2(w-1)}{3}S_0S_3
\nonumber \\
&&~~~~~~~~~~~+\frac{2(w-1)}{3}S_1S_2+6wS_0S_3+2wS_1S_2-8\beta S_0S_2-2\beta S_1^2+2\delta S_0S_1
\bigg{]},
\label{95}
\end{eqnarray}                                 
\begin{eqnarray}
4\lambda S^4&=&4\lambda S_0^4+\frac{16\lambda S_0^3S_1}{r}
+\frac{1}{r^2}\bigg{[}16\lambda S_0^3S_2+24\lambda S_0^2S_1^2\bigg{]}
\nonumber \\
&&+\frac{1}{r^3}\bigg{[}16\lambda S_0^3S_3+48\lambda S_0^2S_1S_2+16\lambda S_0S_1^3\bigg{]}
\nonumber \\
&&+\frac{1}{r^4}\bigg{[}16\lambda S_0^3S_4+24\lambda S_0^2S_2^2+48\lambda S_0^2S_1S_3
+48\lambda S_0S_1^2S_2+4\lambda S_1^4\bigg{]}
\nonumber \\
&&+\frac{1}{r^5}\bigg{[}16\lambda S_0^3S_5+48\lambda S_0^2S_1S_4+48\lambda S_0^2S_2S_3
+48\lambda S_0S_1S_2^2+48\lambda S_0S_1^2S_3
+16\lambda S_1^3S_2\bigg{]},
\nonumber \\
\label{96}
\end{eqnarray}                                 
\begin{eqnarray}
\rho-3p&=&\rho_0-3p_0+\frac{(\rho_1-3p_1)}{r}
+\frac{(\rho_2-3p_2)}{r^2}+\frac{(\rho_3-3p_3)}{r^3}
+\frac{(\rho_4-3p_4)}{r^4}+\frac{(\rho_5-3p_5)}{r^5},
\nonumber \\
\label{97}
\end{eqnarray}                                 
as evaluated in the $r>R$ region. In the above we have carried the expansions as far as needed in order to exhibit at which particular level each of  the explicitly displayed  terms in Eqs. (\ref{93}) and (\ref{94}) first appears. With Eqs. (\ref{95}) - (\ref{97}) all displaying the same power series structure, we see that Eqs. (\ref{82}) and (\ref{83}) will indeed support a power series expansion order by order. The zeroth order term is already satisfied by the background solution given in Eq. (\ref{84}), with the next few orders being maintained by
\begin{eqnarray}
-2kS_0S_1+\gamma S_0^2&=&16\lambda S_0^3S_1 +\frac{(\rho_1-3p_1)}{c},
\nonumber \\
-2kU_0U_1+\gamma U_0^2&=&16\tau U_0^3U_1, 
\nonumber \\
-2kS_0S_2+\gamma S_0S_1+\frac{(w-1)}{3}S_0^2&=&16\lambda S_0^3S_2+24\lambda S_0^2S_1^2 +\frac{(\rho_2-3p_2)}{c},
\nonumber \\
-2kU_0U_2+\gamma U_0U_1+\frac{(w-1)}{3}U_0^2&=&16\tau U_0^3U_2+24\tau U_0^2U_1^2,
\nonumber\\
-4kS_0S_3+2\gamma S_0S_2+\frac{2(w-1)}{3}S_0S_1&=& 16\lambda S_0^3S_3+48\lambda S_0^2S_1S_2+16\lambda S_0S_1^3+\frac{(\rho_3-3p_3)}{c},
\nonumber\\
-4kU_0U_3+2\gamma U_0U_2+\frac{2(w-1)}{3}U_0U_1&=& 16\tau U_0^3U_3+48\tau U_0^2U_1U_2+16\tau U_0U_1^3.
\label{98}
\end{eqnarray}                                 
The pure $U(r)$ sector solution can be simplified to 
\begin{eqnarray}
k&=&-2\tau U_0^2,~~\gamma =12\tau U_0U_1,~~w-1=36\tau U_0U_2+36\tau U_1^2,~~U_3=\frac{U_1^3}{U_0^2},
\nonumber \\
\beta&=&2U_0U_2+72\tau U_0U_1^2U_2 
+72\tau U_0^2U_2^2 +24\tau U_1^4,  
\label{99}
\end{eqnarray} 
with the $S(r)$ sector solution then being satisfied by adjusting the $\rho-3p$ power series terms accordingly.

For the treatment of Eq. (\ref{81}), we note that in it there is a factor of $B(r)$ in the denominator. However, for a de Sitter geometry the coefficient $B(r)$ can actually vanish, with the radial distance $r$ then being bounded by some maximum cut-off value $\Lambda$. (With the large $r$ limit of the candidate $B(r>R)$ metric given in Eq. (\ref{93}) being of the form $B(r>r)\rightarrow w+\gamma r -kr^2$, positivity of the parameters $w$, $\gamma$ and $k$ for instance would entail a cut-off at  $\Lambda \approx (\gamma/2k)[1+(1+4kw/\gamma^2)^{1/2}]$.) For the function $f(r)$ given in Eq. (\ref{81}) to be well-defined, the quantity $\rho+p$ would need to vanish at $r=\Lambda$. Now this quantity actually does vanish at $r=\Lambda$ in the de Sitter background, since as noted in Eq. (\ref{84}), $\rho_0+p_0$ actually vanishes everywhere in the de Sitter background. We thus now need the full $\rho +p$ to vanish at $r=\Lambda$ also. Since the $\rho +p$ and $\rho -3p$ combinations are linearly independent, we are thus free to adjust $\rho +p$ without needing to affect the treatment of Eq. (\ref{82}) which we have just given. To achieve the needed vanishing of $\rho+p$ at $r=\Lambda$ we must set it equal to $(\Lambda -r)D(r)$ where $D(r)$ is a function of $r$ which regular at $r=\Lambda$. (In terms of a $\rho+p$ expansion of the form $\sum (\rho_n+p_n)/r^n$, this is equivalent to setting  $\sum (\rho_n+p_n)/\Lambda^n=0$.) With $B(r)$ also possessing a simple zero at $r=\Lambda$, we can expand the $(\rho+p)/B$ ratio as a power series in $r$ in the $r>R$ region, to obtain
\begin{equation}                                                                              
\frac{3[\rho(r>R)+p(r>R)]}{cB(r>R)}  =R_0+\frac{R_1}{r}+  \frac{R_2}{r^2}+  \frac{R_3}{r^3}+ \frac{R_4}{r^4}+ \frac{R_5}{r^5}.           
\label{100}
\end{equation}                                                  
With the existence of the cut-off, the solution to Eq. (\ref{81}) now takes the form 
\begin{equation}
B(r)=-\frac{r}{2}\int_0^r
dr^{\prime}r^{\prime 2}f(r^{\prime})
-\frac{1}{6r}\int_0^r
dr^{\prime}r^{\prime 4}f(r^{\prime})
-\frac{1}{2}\int_r^\Lambda
dr^{\prime}r^{\prime 3}f(r^{\prime})
-\frac{r^2}{6}\int_r^\Lambda
dr^{\prime}r^{\prime }f(r^{\prime}),
\label{101}
\end{equation}                                 
as now integrated up to $\Lambda$. 

For the candidate expansion of Eq. (\ref{93}) the scalar field term which appears in the source function $f(r)$ in Eq. (\ref{81}) is given by 
\begin{eqnarray}
SS^{\prime \prime}-2S^{\prime 2}&=& \frac{2S_0S_1}{r^3}+ \frac{6S_0S_2}{r^4}+ \frac{12S_0S_3}{r^5}+ \frac{(20S_0S_4+2S_1S_3-2S_2^2)}{r^6} 
\nonumber \\
&&+\frac{(30S_0S_5+6S_1S_4-6S_2S_3)}{r^7},      
\label{102}
\end{eqnarray} 
together with an analogous expression for the $U(r)$-dependent term.  With the expansion of Eq. (\ref{102}) only beginning in order $1/r^3$, we see that there is nothing available to cancel the  $R_0$, $R_1/r$ and $R_2/r^2$ terms which appear in Eq. (\ref{100}). With the insertion of these three particular terms leading to contributions to $B(r>R)$ which respectively behave as $r^4$, $r^3$ and $r^2{\rm log}r$, to recover the candidate expansion of Eq. (\ref{93}), we must thus set $R_0=0$, $R_1=0$, and $R_2=0$. Additionally, we note that  each one of the $1/r^3$, $1/r^4$ and $1/r^5$ terms which appear in Eqs. (\ref{100}) and (\ref{102}) would also generate a term in $B(r)$ which involves  ${\rm log}r$. To recover our candidate form for $B(r>R)$ we must thus cancel all such terms as well, and must thus set 
\begin{equation}                                                                              
-R_3+2S_0S_1+2U_0U_1=0,~~ -R_4+6S_0S_2+6U_0U_2=0,~~ -R_5+12S_0S_3+12U_0U_3=0.          
\label{103}
\end{equation}                                                  
While we need to cancel the contributions of the $1/r^3$, $1/r^4$ and $1/r^5$ terms, for the higher powers everything is well behaved. And not only that, the insertion of such higher powers into Eq. (\ref{101}) will generate back for us the candidate power series for $B(r)$ precisely as given in Eq. (\ref{93}), with the insertion into Eq. (\ref{101}) of  the source term in the generic form $f(r>R)=\sum_6^{\infty}f_n/r^n$ yielding  
\begin{eqnarray}
B(r>R)&=&\sum_{n=6}^{\infty}f_n\bigg{[}-\frac{r}{2(n-3)R^{n-3}}-\frac{1}{6r(n-5)R^{n-5}}
+\frac{1}{2(n-4)\Lambda^{n-4}}+\frac{r^2}{6(n-2)\Lambda^{n-2}} 
\nonumber \\
&&+\frac{1}{r^{n-4}(n-2)(n-3)(n-4)(n-5)}\bigg{]}
\label{104}
\end{eqnarray} 
just as desired. Finally,  to get the full solution, on setting 
\begin{equation}                                                                              
f_{\rm LOC}=  -\frac{3}{4\alpha_g cB}(\rho_{\rm LOC}+p_{\rm LOC}),  
\label{105}
\end{equation}                                                  
we need to augment the globally generated Eq. (\ref{104}) with the additional locally generated
\begin{equation}                                                                              
B(r>R)=-\frac{r}{2}\int_0^R
dr^{\prime}r^{\prime 2}f_{\rm LOC}(r^{\prime})
-\frac{1}{6r}\int_0^R
dr^{\prime}r^{\prime 4}f_{\rm LOC}(r^{\prime}),
\label{106}
\end{equation}                                                  
to thus yield a net $1/r$ term in $B(r>R)$ of the form
\begin{equation}                                                                              
B(r>R)=-\frac{1}{6r}\int_0^R
dr^{\prime}r^{\prime 4}f_{\rm LOC}(r^{\prime})-\sum_{n=6}^{\infty}\frac{f_n}{6r(n-5)R^{n-5}}.
\label{107}
\end{equation}                                                  
With regard to Eq. (\ref{107}), we note that even though it is difficult to assess the relative importance of the various terms which appear in it, with the entire source term of Eq. (\ref{81}) being multiplied by an overall $1/\alpha_g$ factor, we can nonetheless conclude that there will be some choice for the sign of $\alpha_g$ for which the net effect in Eq. (\ref{107}) will be attractive just as desired, with the signs of the various terms which appear in Eqs. (\ref{98}) and (\ref{99}) adjusting accordingly. As we thus see, an attractive $1/r$ potential can naturally arise in conformal gravity even when long range macrosopic scalar fields are present.

\section{Modification of trajectories due to the scalar field} 

With the scalar field modifications to the exterior metric associated with Eqs. (\ref{104}) and (\ref{106}) beginning in order $1/r^2$, a precise enough monitoring of the geometry in the vicinity of the sun could thus reveal the presence of any macroscopic scalar field that the sun might possess. For any monitoring which involves photons, we note that since massless particles do not couple to the mass generating $S(r)$, for light rays one can continue to use the standard massless test particle geodesics, as evaluated now in the metric associated with Eqs. (\ref{104}) and (\ref{106}). For massive particles however, there is an additional effect, since their very coupling to the spatially dependent $S(r)$ itself also modifies their motion. To study this effect in detail we would need to eikonalize the Dirac wave equation given in Eq. (\ref{3}) to find the trajectories of its short wavelength ray solutions. To get a sense as to what such eikonal trajectories might look like, we can instead vary the test particle action
\begin{equation}                                                                              
I_T=-h\int d\tau S(x), 
\label{108}
\end{equation}                                                  
as this action not only reduces to the conventionally used massive test particle action when $S(x)$ is constant, but for varying $S(x)$ it is actually fully conformal invariant, since the $e^{-\alpha(x)}$ change in the scalar field is compensated by an accompanying $e^{\alpha(x)}$ change in the proper time. Variation of the action of Eq. (\ref{108}) with respect to the coordinates of the test particle thus yields conformal invariant trajectories, with their specific form being given as \cite{Mannheim1993}
\begin{equation}                                                                              
hS\left(
\frac{d^2x^{\lambda} }{ d\tau^2} +\Gamma^{\lambda}_{\mu \nu} 
\frac{dx^{\mu}}{d\tau}\frac{dx^{\nu } }{ d\tau} \right) 
= -hS_{;\beta} \left( g^{\lambda
\beta}+
\frac{dx^{\lambda}}{d\tau}                                                      
\frac{dx^{\beta}}{d\tau}\right).
\label{109}
\end{equation}                                                  
With the Yukawa coupling constant $h$ canceling in Eq. (\ref{109}), and with all fundamental fermions getting their masses from one and the same $SU(2)\times U(1)$ Higgs doublet (with masses that only differ because each fermion has its own associated $h$), we see that even with a spatially varying $S(r)$, all particles will nonetheless still fall in a gravitational field with trajectories which are independent of their masses.

To determine the specific trajectories which are involved in the static, spherically symmetric source case of interest, we note that for the general standard coordinate metric given in Eq. (\ref{39}), the four equations of motion contained in Eq. (\ref{109}) take the form
\begin{eqnarray}
&&\frac{d^2t}{d\tau^2}+\frac{B^{\prime}}{B}\frac{dt}{d\tau}\frac{dr}{d\tau}
=-\frac{S^{\prime}}{S}\frac{dt}{d\tau}\frac{dr}{d\tau},
\nonumber \\
&&\frac{d^2r}{d\tau^2}+\frac{A^{\prime}}{2A}\left(\frac{dr}{d\tau}\right)^2
-\frac{r}{A}\left(\frac{d\theta}{d\tau}\right)^2
- \frac{r{\rm sin}^2\theta}{A}\left(\frac{d\phi}{d\tau}\right)^2
+\frac{B^{\prime}}{2A}\left(\frac{dt}{d\tau}\right)^2
=-\frac{S^{\prime}}{AS} 
-\frac{S^{\prime}}{S}\left(\frac{dr}{d\tau}\right)^2
\nonumber \\
&&\frac{d^2\theta}{d\tau^2}
+\frac{2}{r}\frac{d\theta}{d\tau}\frac{dr}{d\tau}
-{\rm sin}\theta{\rm cos}\theta\left(\frac{d\phi}{d\tau}\right)^2
=-\frac{S^{\prime}}{S}\frac{d\theta}{d\tau}\frac{dr}{d\tau},
\nonumber \\
&&\frac{d^2\phi}{d\tau^2}
+\frac{2}{r}\frac{d\phi}{d\tau}\frac{dr}{d\tau}
+2{\rm cot}\theta\frac{d\phi}{d\tau}\frac{d\theta}{d\tau}
=-\frac{S^{\prime}}{S}\frac{d\phi}{d\tau}\frac{dr}{d\tau}.
\label{110}
\end{eqnarray}                                 
As we thus see, even in the presence of  a spatially dependent scalar field, the equations of motion still admit of trajectories with fixed $\theta=\pi/2$. In such trajectories we find that the three other equations of motion admit of exact first integrals
\begin{eqnarray}
BS\frac{dt}{d\tau}&=&C,
\nonumber \\
S^2+AS^2\left(\frac{dr}{d\tau}\right)^2
+\frac{K^2}{r^2}-\frac{C^2}{B}&=&D,
\nonumber \\
r^2S\frac{d\phi}{d\tau}&=&K,
\label{111}
\end{eqnarray}                                 
where $C$, $D$ and $K$ are integration constants. Finally, on eliminating the dependence on $d\tau$, we find that the trajectories can be written as
\begin{eqnarray}
S^2+\frac{AC^2}{B^2}\left(\frac{dr}{dt}\right)^2
+\frac{K}{r^2}-\frac{C^2}{B}&=&D,
\nonumber \\
S^2+\frac{AK^2}{r^4}\left(\frac{dr}{d\phi}\right)^2
+\frac{K^2}{r^2}-\frac{C^2}{B^2}&=&D,
\nonumber \\
\frac{r^2}{B}\frac{d\phi}{dt}&=&\frac{K}{C}.
\label{112}
\end{eqnarray}                                 
When $S$ is constant these relations all reduce to the standard massive particle trajectories, with the modifications when $S$ is spatially varying then providing a possible window on symmetry breaking physics as well as constraints on the magnitudes of such modifications.

The author would like to acknowledge helpful conversations with Dr. E. E. Flanagan, Dr. K. Van Acoleyen, Dr. W. Moreau and J. Poveromo.

\end{document}